# Incised-valley morphologies and sedimentary-fills within the inner shelf of the Bay of Biscay (France): a synthesis

Eric Chaumillon[1], Jean-Noël Proust[2], David Menier[3] and Nicolas Weber[4]

[1] Centre Littoral De Géophysique, Université de La Rochelle, Pôle Sciences, Avenue Michel Crépeau, 17042 La Rochelle, Cedex 01, France, - echaumil@univ-lr.fr

[2] UMR6118 CNRS Géosciences-Rennes, Université de Rennes1, Campus de Beaulieu, 35042 Rennes Cedex, France

[3] LEMEL, Université Bretagne Sud, Rue André Lwoff, 56017 Vannes Cedex, France

[4] EPSHOM - Section Géodésie-Géophysique - Bureau des Ouvrages et des Produits - 13, rue du Chatelier, BP 30316, 29603, Brest Cedex, France

**Abstract**

This study is a first synthesis focused on incised-valleys located within the inner shelf of the Bay of Biscay. It is based on previously published results obtained during recent seismic surveys and coring campaigns. The morphology of the valleys appears to strongly controlled by tectonics and lithology. The Pleistocene sedimentary cover of the shelf is very thin and discontinuous with a maximum thickness ranging between 30 and 40 m in incised valley fills. Thus the incised bedrock morphology plays a key-role by controlling hydrodynamics and related sediment transport and deposition that explains some variations of those incised-valley fills with respect to the previously published general models.

## 1. Introduction

Incised-valleys are key-areas for the sedimentary record of sea level variations on continental margins (Zaitlin et al., 1994). Their incision pattern records decrease in accommodation rates and the bulk of their infill mainly corresponds to sediments emplaced during marine transgression. Incised-valleys display two major kinds of sediment fills (Zaitlin et al., 1994): simple fills corresponding to a single depositional sequence (Lericolais et al., 2001, Li et al., 2002, Gutierez et al., 2003, Weber et al., 2004a and b) and compound fills corresponding to multiple, superimposed, cycles of incision and deposition (Thomas and Anderson, 1994, Foyle and Oertel, 1997, Proust et al, 2001, Tesson, et al., 2005).

Within the Bay of Biscay, the pioneer work on incised-valleys corresponds to the study of the Gironde estuary (Allen and Posamentier, 1993, 1994) which becomes a reference model for mixed tide- and wave-dominated incised-valleys. The following work of Lericolais et al. (2001) showed that high resolution seismic profiling brought new insights on the seaward pinching out of the Gironde incised-valley. Since 2000, several new high and very high resolution seismic surveys have been conducted on drowned incised-valleys of the inner shelf of the northern and eastern Bay of Biscay. From south to north, those studied incised valleys include : (1) the Leyre incised-valley (Féniès and Lericolais, 2005), (2) the Seudre incised-valley (Bertin et al., 2004), (3) the Charente incised-valley (Chaumillon et al., 2004, Weber et al., 2004a, Weber, 2004, Chaumillon and Weber, 2006), (4) the Lay-Sèvre incised-valley (Chaumillon et al., 2004, Weber et al., 2004b, Weber, 2004, Chaumillon and Weber, 2006), (5) the Loire incised-valley (Chaumillon and Proust, 2005; Renault, 2006), (6) the Artimon incised-valley (Menier et al., 2006), (7) the Vilaine incised-valley (Proust et al., 2001, Menier, 2004, Menier et al., 2006), (8) The Etel incised-valley (Menier, 2004, Menier et al., 2006), (9) the Lorient incised-valley (Menier, 2004, Menier et al., 2006) and (10) the Concarneau incised-valley (Loget, 2001; Menier, 2004, Menier et al., 2006).

This paper is the first attempt to summarize, at the scale of the Bay of Biscay, the interpretation of the recent, dense grid of seismic profiles acquired on the incised valley network. The aim is to compare the morphologies and sedimentary-fills of the inner shelf incised-valleys to the reference models (Zaitlin et al., 1994, Ashley and Sheridan, 1994, Allen and Posamentier, 1994). These models present the overall stratigraphy of the sediments deposited in an incised valley during a single cycle of sea-level variation and document controlling factors such as the valley shape, the relative influence of waves and tides, the depositional gradient, the sea-level changes, the sediment supply and the topographic relief of the valley (i.e. preservation potential). However, it appears difficult to differentiate the respective influence of each factor. One way to explore this issue is to compare valleys that share same depositional conditions and in the same basin. The comparison of the different incised-valleys of the Bay of Biscay aimed to illustrate the large variability of the nature of their infillings with respect to the reference models and to decipher among the numerous driving processes responsible for their variability.

## 2. Studied area

The French Atlantic coast of the Bay of Biscay belongs to a passive continental margin (Montadert et al., 1971, Thinon et al., 2001). This margin is inherited from the opening of the Bay of Biscay and an anti-clockwise rotation of the Iberian Peninsula during the Cretaceous (Stauffer and Tarling, 1971). Major faults of this margin are oriented in a northwest-southeast direction, which is parallel to the structural pattern inherited from reactivated Hercynian faults. Since the opening of the Bay of Biscay fault reactivations occurred in response to the Pyrenean and Alpine collisions. Present-day seismic activity in the western part of France is moderate (Müller et al., 1992). It is possibly related to strain originating at the southern European plate boundary (Ziegler, 1992).

Onland, rocky outcrops correspond to two main sets: (1) metamorphic and magmatic rocks related to the Hercynian orogen, north of the lay-Sèvre estuary (Pertuis breton, Fig. 1) and (2) Mesozoic and Cenozoic sedimentary rocks including, from north to south, early Jurassic to Pliocene strata in the Aquitanian basin, south of the Lay-Sèvre estuary. Southward of the Gironde estuary, Cenozoic strata are overlain by an extensive Quaternary sand cover including large eolian sand dunes.

The northern and eastern coastlines of the Bay of Biscay are interrupted by major embayments, including, from south to north (Fig. 1): the Leyre Estuary (Féniès and Lericolais, 2005), the Gironde Estuary (Allen and Posamentier, 1994), the Pertuis d'Antioche and the Marennes-Oléron Bay (Bertin et al., 2005, Chaumillon and Weber, 2006, Weber et al., 2004a,), the Pertuis Breton (Weber et al., 2004a, Chaumillon and Weber, 2006), the Loire Estuary, the Vilaine estuary, the Quiberon and the Morbihan Bays (Proust et al., 2001). Westward of the Quiberon Bay, the south coast of Brittany shows small estuaries or rias (Fig. 1), including from east to west, The Etel, the Blavet, the Laïta and the Odet estuaries (Loget, 2001; Menier, 2004, Menier et al., 2006). The age of the down cutting of those estuaries and related valley segments is still a matter of debate. They are dated either Late Pleistocene (120-18kyrs BP) (Fairbanks, 1989, Prell et al., 1986) or from previous lowstands of sea level (Chaumillon and Weber, 2006, Féniès and Lericolais, 2005, Lericolais et al., 2001, Menier et al., 2006, Proust et al., 2001, Weber et al., 2004a, 2004b). The age of the transgressive infilling of the valleys, is less controversial. It occurred probably from 18 to 6kyrs BP although some older ages are mentioned locally (Proust et al., 2001; Laurent, 1993; Van Vliet-Lanoë et al., 1995, 1997). During the sea-level rise, the incised valleys where flooded and transformed into estuaries.

Present-day rivers connected to estuaries in the Bay of Biscay (Fig. 1) belong to three main categories: (1) large rivers (drainage basin area > 50000 km$^2$, water discharge > 500 m$^3$.s$^{-1}$, Table 1): the Garonne and Dordogne (Gironde estuary) and the Loire rivers, (2) smalls rivers (500 km$^2$ < drainage basin area < 50000 km$^2$, 10 m$^3$.s$^{-1}$ < water discharge < 500 m$^3$.s$^{-1}$, Table 1): the Charente, Lay-Sèvre, Vilaine and Blavet rivers and (3) very small rivers (drainage basin < 500 km$^2$): the Leyre, Seudre, Etel, Laïta, and Odet. The Gironde and Loire are the main sources for fine sediments exported from the continent to the shelf (Allen, 1972, Castaing and Jouaneau, 1987, Figueres et al., 1985).

Tides in the Bay of Biscay are semi-diurnal. Mean spring tide range varies from less than 4 m (Leyre Estuary) to more than 5 m (Charente and Loire Estuaries). In the Gironde mouth and the Vilaine, Etel, Laïta, Belon and Aven estuaries, the average spring tide range is between 4 and 5 m. The yearly average significant wave height in the Bay of Biscay is about 1.5 m, whereas wave height during storm events can exceed 6 m (Barthe and Castaing, 1989). Swells are predominantly coming from west to north-west inducing a southward net littoral drift in the eastern part of the Bay of Biscay. With regards to these hydrodynamic parameters, the estuaries of the Bay of Biscay display both tide and mixed-tide and wave-dominated environments and the Bay of Biscay can be considered as a wave dominated shelf.

The paleovalley network of the inner shelf of the Bay of Biscay and the deposits above the incision surface was partly known for three decades (Boillot et al., 1971; Vanney, 1977). But due to the lack of resolution of earlier seismic data and to the wide spacing of seismic lines, these data could not provide detailed information on the detailed morphology and the geometry of sediments filling the studied valleys.

## 3. Methods

This study is mostly based on the interpretation of recently published seismic data (Fig. 2 and 3) acquired with Sparker sources (band pass frequencies ranging from 200 to 1200 Hz, table 2) associated with mono channel streamers (Chaumillon et al., 2004, Féniès and Lericolais, 2005, Lericolais et al., 2001, Menier, 2004, Proust et al., 2001, Weber, 2004, Weber et al., 2004a, 2004b). All seismic data were recorded digitally. After seismic processing, interpretation of seismic profiles was run following the principles of seismic stratigraphy (Mitchum et al., 1977, Payton, 1977). Lithological groundtruthing of seismic unit interpretation was provided by cored drill holes (Vilaine incised-valley, Proust et al., 2001, Gironde incised-valley, Lericolais et al., 2001 and Leyre incised-valley, Féniès and Lericolais, 2005) or vibrocores (Charente and lay-Sèvre incised-valleys, Weber et al., 2004a, 2004b, Chaumillon and Weber, 2006) and allowed to collect sediments ranging from silt and clay to sand and gravels. For P-wave travel time to depth correction and the production of isopach maps of the valley–filling sediments (Fig. 4), we used a P wave velocity of 1500 m.s$^{-1}$ in the water body and 1600 to 1800 m.s$^{-1}$ in unconsolidated sediment, according to the relationships between grain size of the collected sediment (mainly silts and sands, Chaumillon et al., 2004, Chaumillon and Weber, 2006, Féniès and Lericolais, 2005, Lericolais et al., 2001, Proust et al., 2001, Weber et al., 2004a, 2004b) and P-wave velocities (Hamilton, 1972). Radiocarbon ages were obtained on well preserved, full-bodied shells (Chaumillon et al., 2004, Chaumillon and Weber, 2006, Menier et al., 2006, Proust et al., 2001).

# 4. Results

Previously published seismic results (Chaumillon and Weber, 2006, Lericolais et al, 2001, Menier et al, 2006, Proust et al, 2001, Weber et al, 2004a) were used to make a contour map of the incised-valleys revealed by seismic investigations within the inner shelf of the northern and eastern Bay of Biscay (Fig. 1). The major variations in valley morphologies and valley-fills are evidenced thank to a selection of across strike seismic lines including: (1) interpreted seismic lines crossing the eight main incised-valleys of the northern and eastern Bay of Biscay  (Fig. 2) ; (2) seismic profiles showing the three typical class of valley-fills observed within the Bay of Biscay (Fig. 3). The choice of the seismic line corresponds to the most typical profile illustrating the valley-fill among previously published profiles with the exception of the Loire incised-valley which is illustrated by a new seismic profile. Most of the seismic units are correlated with well cores or vibrocores (Chaumillon et al., 2004, Chaumillon and Weber, 2006, Féniès and Lericolais, 2005, Lericolais et al., 2001, Proust et al., 2001, Weber et al., 2004a, 2004b). Valley shape and valley-fill similarities and differences between the studied incised-valleys are presented respectively in the following.

## 4.1. Valley shape

On the whole studied area, seismic profiles exhibit a major regional erosion surface, at the top of rocky basement acoustic units (Fig. 3). This regional surface forms the base of different valleys that are incised into the bedrock (Fig. 2 and 3). This surface is interpreted as an Type 2 sequence boundary (Jervey, 1988) probably formed during Late Cenozoic lowstands and strongly reworked recently, during the last 20-18kyrs-old sea-level drop and subsequent lowstand (Chaumillon and  Weber, 2006, Féniès and Lericolais, 2005, Lericolais et al., 2001, Menier et al., 2006, Proust et al., 2001, Weber et al., 2004a, 2004b).

The contour map of incised-valleys (Fig. 1) clearly evidences their connection with present-day rivers. Two kinds of channel networks are observed: (1) linear longitudinal networks (Etel, Concarneau and Lorient incised-valleys) when valleys follow NW-SE faults scarps and (2) rectangular networks (Vilaine, Lay-Sèvre,

Charente incised valleys) when valleys are successively parallel and oblique to the faults. Thus the orientations of the valley incisions appear strongly controlled by the tectonic heritage through neo tectonic reactivations.

Some paleovalleys deeply incise soft marly sediment and parallel the hard carbonate strata. These marl-rich strata are of Late Kimmeridgian age for the Charente valley and Callovian to Early Oxfordian for the Lay-Sèvre valley (Weber, 2004). These observations demonstrate the role of bedrock lithology as a control of the orientations of valley incisions. The Charente and Lay-Sèvre valleys show many examples within the Pertuis d'Antioche and Pertuis Breton (Fig. 1 and 4),

The contour map of the main incised-valleys shows that the incisions crossing the shelf are discontinuous (Fig. 1). The main submarine rocky outcrops that correspond to the interfluves of most of the valleys are widespread north of the Gironde Valley (Fig. 1). The valley incisions exhibit a progressive decrease and wedging out of incision depths in a seaward direction. This decrease of incision depths occurs below 40 to 60 m below sea level, where the shelf gradient decreases (Cirac et al., 2000), indicating that the seaward termination of the incision is related to the shelf morphology. This phenomenon is also observed elsewhere e.g. on the Central Texas Shelf (Eckles et al., 2004).

## 4.2. Valley-fill similarities and differences

Valley-fill thicknesses usually range from 30 to 40 ms two way travel time (Fig. 2 and 3), that is about 25 to 35 m. Netherthelesss, in some places, sediment thickness considerably decreases leading to a discontinuous valley-fills and disconnected depocenters within the valley-fills (Fig. 4). The areas where valley-fills taper or disappear corresponds to present-day estuary mouth of the Gironde (Berné et al., 1993) or entrenched parts of the incised-valleys, e.g. the Charente, Lay-Sèvre, valleys and south-Brittany valleys (Chaumillon and Weber, 2006, Menier et al., 2006) (Fig. 4). As sandwaves are commonly observed within those areas (Berné et al. 1993, Weber and Chaumillon, 2004, Chaumillon and Weber, 2006), tidal scouring caused by strong tidal currents, related to the entrenched morphology of the valleys or the estuary mouth, have probably played a role by removing sediments and/or preventing accumulation from those valley segments (Chaumillon and Weber, 2006).

Valley-fills generally include 4 main seismic units (Fig. 2 and 3, Table 3, Chaumillon and Proust, 2005). From base to top, they are the followings.

(1) U1 (S1 in Lericolais et al., 2001, $U_{ios}1$, $U_{bos}1$, in Chaumillon and Weber, 2006, unit 4 in Proust et al., 2001, unit 1 in Menier et al., 2006) is composed of distinct seismic sub-units of small lateral extent that are isolated from each other in the case of valleys connected to small rivers (Concarneau, Lorient, Vilaine, Lay-Sèvre, Charente and Leyre examples, Fig. 2 and 3). U1 corresponds to large and relatively thick units in the cases of valleys connected to large rivers (Loire and Gironde examples, Fig. 2). These sub-units show internal inclined reflectors. U1 was interpreted as fluvial sands and gravels belonging to the lowstand system tract (Chaumillon and Weber, 2006, Lericolais et al., 2001, Menier et al., 2006, Proust et al., 2001).

(2) U2 (S2 in Lericolais et al., 2001, $U_{ios}2$, in Chaumillon and Weber, 2006, unit 5 in Proust et al., 2001, unit 3 in Menier et al., 2006) is a large and thick unit. It generally includes sub horizontal reflectors and internal channelised unconformities. U2 was interpreted as estuarine mud or mixed sand and mud belonging to the transgressive system tract (Chaumillon and Weber, 2006, Lericolais et al., 2001, Menier et al., 2006).

(3) U3 (S3 in Lericolais et al., 2001, $U_{ios}3$, $U_{bos}2$, $U_{bid}1$, in Chaumillon and Weber, 2006, US1 in Féniès and Lericolais, 2005, unit 6 in Proust et al., 2001, unit 4 in Menier et al., 2006) is a large and thick unit. It generally includes inclined reflectors and internal channelised unconformities. U3 was interpreted as estuary mouth and tidal channel coarse sands belonging to the transgressive system tract (Chaumillon and Weber, 2006, Féniès and Lericolais, 2005, Lericolais et al., 2001, Menier et al., 2006, Proust et al., 2001). U2 and U3 units correspond to the bulk of the different valley-fills.

(4) U4 (S4 in Lericolais et al., 2001, $U_{ios}4$, $U_{bos}3$, $U_{aid}4_{hac}$, $U_{aid}4_{sd}$, $U_{bid}2$, in Chaumillon and Weber, 2006, unit 6 in Proust et al., 2001, unit 5 in Menier et al., 2006) generally consists of a sheet drape including sub parallel and sub horizontal reflectors or sandbanks containing inclined reflectors (Chaumillon et al., 2002, Chaumillon et al., 2004). U4 upper surface is the present-day seafloor. U4 sheet drape is made of mud (Pertuis d'Antioche, Pertuis Breton, Vilaine, Concarneau and Lorient incised-valleys) or shoreface sands (outer shoals of the Charente and Lay-Sèvre, Leyre incised-valleys) or shoreface mixed sand and mud (Gironde incised-valley). U4 was interpreted as the highstand system tract (Chaumillon and Weber, 2006, Lericolais et al., 2001, Menier et al., 2006, Proust et al., 2001) or the late transgressive system tract reworked by present-day hydrodynamics, in the offshore part of the Charente and Lay-Sèvre incised valleys (Chaumillon and Weber, 2006)..

Two valley-fills show important differences with reference to this 4 fold valley-fill: the Leyre and the Lay-Sèvre incised-valleys (within the Pertuis Breton). Féniès and Lericolais (2005) have proposed that the Leyre incised-valley is entirely filled by the transgressive system tract and that the lowstand system tract is absent. The small progradational unit resting at the base of Leyre valley is very similar to seismic units U1 of the Concarneau, Vilaine, Loire and Charente incised-valleys (Fig. 2) and could correspond to sands and gravels belonging to the lowstand system tract. Whatever the interpretation, both the absence (Féniès and Lericolais, 2005) or the small size of the lowstand system tract (our interpretation), may be due to the depth of fluvial incision during sea-level fall shallower than the depth of the tidal ravinement surface generated during the subsequent rise of sea-level (Féniès and Lericolais, 2005). The highstand system tract of the Leyre incised-valley is reduced to a thin sand drape resting over a wave ravinement surface at the top of the transgressive system tract. This is similar to the sand drapes observed in the offshore part of the Charente and Lay-Sèvre incised-valleys (inter island and Breton outer shoal, Chaumillon and Weber, 2006) where the highstand system tract was assumed to be absent (Fig. 2, Weber et al., 2004a, Chaumillon and Weber, 2006). An alternative hypothesis may consider those sand drapes made up of reworked transgressive sands and belonging to a much reduced highstand system tract. Whatever the interpretation, the top of the valley-fill successions, of the Leyre, Charente and Lay-Sèvre, belongs to the shoreface domain, undergoing sediment reworking and transport driven by swell orbital motions (Idier et al., 2006). The Lay-Sèvre valley-fill (Fig. 2 and 3) mainly consists of a succession of seismic units showing high-angle inclined reflectors composed of marine sand (Weber et al., 2004b) with internal channelized unconformities (U3 or $U_{bid}1$, in Chaumillon and Weber, 2006). This sandy succession is interpreted as sandbanks, sandspits or tidal-delta deposits, emplaced under the control of tides and waves. Those stacked sandbodies are covered by a thin drape (U4 in this paper or $U_{bid}2$, in Chaumillon and Weber, 2006), made up of recent (about 1000 years BP, table 3) tidal estuarine mud (Billeaud et al., 2005, Chaumillon et al., 2004, Chaumillon and Weber, 2006, Weber et al., 2004b).

The upper successions of all the studied valleys-fills generally show a fining upward trend. It corresponds to the transition from transgressive coarse sands (U3 unit) to highstand mud (Pertuis d'Antioche, Pertuis Breton, Vilaine, Concarneau and Lorient incised-valleys) or tidal sandbank (Pertuis d'Antioche) or shoreface sands (outer shoal of the Charente, Leyre incised-valleys) or shoreface mixed sand and mud (Gironde incised-valley).

The main differences between the valleys of the northern and eastern Bay of Biscay correspond to the nature of their sedimentary-fills. Correlations between seismic units and cores show that low angle to sub horizontal internal reflectors are generally correlated to mud-dominated units as seismic units displaying high angle clinoforms are correlated to sand-dominated units (Billeaud et al., 2005; Chaumillon et al., 2002, 2004, Féniès and Lericolais, 2001, Lericolais et al., 2001, Proust et al., 2001, Weber et al., 2004a, 2004b). Based on these observations, three main kinds of valley-fills are distinguished:

(1)    Valley-fills displaying alternations of high angle reflector sandy units (generally U1 and U3) and sub horizontal to low-angle reflector fine sand to mud-rich units (generally U2 and eventually U4). This alternation has been named the "seismic sandwich" (Weber et al., 2004a) in reference to the Ashley and Sheridan "Sedimentary Sandwich" (Ashley and Sheridan, 1994). Those valley-fills include the Charente outer shoal (Weber et al., 2004a, Chaumillon and Weber, 2006), the Etel (Menier et al., 2006), the Loire (Renault, 2006) and the Gironde (Lericolais et al., 2001) incised-valleys.

(2)    Valleys-fills mainly displaying high-angle progradational sandy units with tidal ravinement surfaces. Those valleys correspond to the Leyre (Féniès and Lericolais, 2005) and the Pertuis Breton (Weber et al., 2004b, Chaumillon and Weber, 2006) incised-valleys.

(3)    Valley-fills displaying thick sub horizontal aggrading muddy units. Those valleys include the Concarneau (Loget, 2001), Lorient Menier et al., 2006), Vilaine (Proust et al., 2001) and the Pertuis d'Antioche (Chaumillon and Weber, 2006) incised-valleys.

In terms of sequence stratigraphy, single fifth order sequence seems to be the dominant feature (Charente, Etel, Leyre, Lay-Sèvre, Gironde, Concarneau and Lorient Valleys) but the Vilaine Valley-fill may include two sequences. The age of the basal sequence remains unknown.

## 5. Discussion

**Comparison of the incised-valleys of the northern and eastern Bay of Biscay with previously published models**

The northern and eastern Bay of Biscay exhibit a large variability in the nature of the valley-fills from sand-dominated valley-fills (Leyre and Lay-Sèvre incised-valleys), to mixed sand and mud valley-fills (Charente outer shoal, Etel, Loire and Gironde incised-valleys) and mud-dominated valley-fills (Lorient, Concarneau, Vilaine and Pertuis d'Antioche incised-valleys). This variability is not fully illustrated by the general valley-fill model (Zaitlin et al., 1994) or the model proposed for the Eastern Atlantic margin of the United States (Ashley and Sheridan, 1994).

The studied valleys of the Bay of Biscay are located in the same basin and thus should have experienced similar relative sea level and climate changes. The influence of relative sea level rise explains the observed four-fold valley-fill successions and also the fining upward trend and drop in hydrodynamic energy of the upper successions of the valley-fills through the transition from the Holocene transgression to present-day sea-level highstand.

However, relative sea level and climate changes cannot explain all of the observed differences between the incised-valleys of the northern and eastern Bay of Biscay such as the thickness and volume of seismic units and the ratio between sand and mud within the valley-fills. Other forcing parameters may explain these variations like hydrodynamics (waves, tides, river water discharges), sediment supply, valley shape or the size of the drainage area of the rivers. In the later case, we observe that the U1 lowstand basal units consist of thin seismic packages in the case of valleys connected to small rivers (Leyre, Charente, Lay-Sèvre, Vilaine, Blavet, Odet), and U1 corresponds to large seismic units when valleys are connected to large rivers (Gironde and Loire). This would indicate that sedimentation, during lowstand periods, was much more important in large rivers than in small rivers.

On the other hand, these rivers exhibit incised valley fills of very different natures, from mud to sand or mixed sand and mud-dominated successions, that experienced similar preservation history and hydrodynamic conditions. Available cores (Féniès and Lericolais, 2005, Lericolais et al., 2001, Proust et al., 2001, Weber et al., 2004a and 2004b) show that those valley-fills are mainly made up of marine sediments, whose variations, from sand to mud-dominated valley-fills, could be mainly related to marine processes and marine sediment supply as required for the infillings of most estuaries (Fig. 5).

The mixed (sand and mud) incised-valleys (Charente, Etel, Loire and Gironde incised-valleys) characterised by their high angle–low angle–high angle internal reflection pattern succession which represents the "Seismic Sandwich" (Weber et al., 2004a) give a seismic validation for the model of large valley fill proposed by Ashley & Sheridan (1994) and for the model of mixed, tide- and wave-dominated valley-fill (Allen and Posamentier, 1994). However, other valley-fills like our sand-dominated and the mud-dominated end-members differ from those models.

The sand-dominated valleys include the Leyre and the Lay-Sèvre incised-valleys. The Leyre incised-valley is mainly filled by the transgressive system tract, made of shelly coarse-grained sands and gravels (Féniès and Lericolais, 2005). The lowstand and highstand system tracts are reduced or absent (Fig. 2). In this example the abundance of sand can be explained by hydrodynamic processes that have removed sediments of the lowstand and highstand system tracts and have led to a valley-fill dominated by high energy tidal channel deposits (Féniès and Lericolais, 2005). Two processes of sediment removal have successively occurred during transgression: (1) tidal ravinement, when the valley became an estuary and (2) wave ravinement, when the estuary was drowned. Sediment removal during transgression is mainly controlled by the depth of erosion (Ashley and Sheridan, 1994), itself controlled by the wave climate. In the case of the Leyre incised-valley, located in a storm-dominated open-shelf, high-energy wave climate are assumed to explain the large amount of erosion. The sand-dominated Lay-Sèvre valley-fill mainly shows a succession of high-angle clinoforms (high angle reflectors) composed of marine sand covered by a thin mud drape (Fig. 2 and 3, Chaumillon and Weber, 2006, Weber et al., 2004b). This valley-fill, composed of sandbanks, sandspits and tidal-delta deposits, contrasts with the inner domain (Pertuis d'Antioche) of the Charente valley-fill, located immediately to the south, which is mud to mixed sand and mud-dominated (Chaumillon et al., 2006, Weber et al., 2004a). Taking into account the morphological similarities between those two valleys that are close to each other and the marine facies of the two valley-fills (Chaumillon and Weber, 2006), the observed differences in sediment fill may result from differences in the marine sediment supply. The northern valley (Lay-Sèvre incised-valley), located updrift with respect to the southern valley (Charente incised-valley), is a sediment trap for sand transported by the southward-moving littoral drift (Chaumillon and Weber, 2006). Thus, valley location with respect to littoral drift seems to be a critical parameter that controls internal architecture and sedimentary facies of incised-valley fills. In addition the lack, in the Leyre valley-fill, or the low amount of mud, in the Lay-Sèvre valley-fill, may be related to their locations, far from the Gironde and Loire estuaries which are the main source for suspended matter (Allen, 1972, Castaing and Jouaneau, 1987, Figueres et al., 1985).

The mud-dominated incised-valleys include the Concarneau, Lorient, Vilaine and inner domain of the Charente incised-valleys. The Concarneau and lorient are narrow and linear valleys. A large part of those valley-fills consist of the U2 muddy estuarine unit (unit 3 in Menier et al., 2006). This abundance of mud is explained by the size of the tidal prism which is controlled by the valley shapes when the valleys were drowned and became estuaries. In those cases of narrow and linear valleys that lack tributaries, the tidal prism is smaller than in the case of large and dendritic valleys (Vilaine valley example, Menier et al., 2006). As tidal currents are scaled to the ratio of the tidal prism to the entire volume of the incised-valley, tidal currents are weaker in the case of narrow and linear valleys with respect to large and dendritic valleys, leading to higher deposition of mud during transgression. The higher amount of mud in narrow valleys can also be explained by the tidal flow damping. Hence, those linear valleys are isolated from the open sea by narrow passes where tidal flow is damped by friction on the valley walls and tidal wave energy is dissipated by diffraction.

The Vilaine mud-dominated valley differs from the Concarneau and Lorient valleys because of the abundance of mud, related to the thick uppermost U4 unit which corresponds to highstand offshore muds (Proust et al., 2001, Menier et al., 2006). Since the last sea-level still stand (about 6000 years BP), the Vilaine valley is located in a semi-enclosed tide-dominated environment where the shielding action of rocky islands and headlands stops littoral drift and attenuates onshore sand transport by waves. This semi-enclosed environment, due to the bedrock morphology, would be responsible for deposition of high amount of mud during the highstand period.

Another character of the valleys of the northern and eastern Bay of Biscay, which is not illustrated by the published valley-fill model (Zaitlin et al., 1994, Ashley and Sheridan, 1994), corresponds to the lack of continuity of their infilling (Fig. 4). This can be explained by large sediment volumes removal related to tidal scouring due to a wide mean tidal range (Chaumillon and Weber, 2006, Liu et al., 1998) enhanced by local coastline convergence. Discontinuous valley-fills are also related to the thin soft sedimentary cover within the inner shelf (Fig. 1), this sediment starvation is explained by low sediment input from the land and combination of low subsidence and frequent and high amplitude sea-level changes during the Quaternary leading to strong sediment reworking.

From this discussion, it appears that both the great variability of valley-fills of the northern and eastern Bay of Biscay (Fig. 2, 3 and 5) and the occurrence of discontinuous valley-fills (Fig. 4) can be explained by variations in marine hydrodynamics (wave and tides) and sediment supply. In most of the studied examples, variations in hydrodynamics (swell attenuation for the Vilaine incised-valley, attenuation of tidal prism and/or of tide-induced currents in the cases of the Concarneau and Lorient incised-valleys, enhancement of tide-induced currents due to local coastline convergence) and variations in marine sediment supply (updrift valley acting as a sediment trap in the case of the Lay-Sèvre incised-valley) are related to the shape of the valley and the morphology of the bedrock. Valley-fills of the Bay of Biscay, bounded by rocky outcrops and connected to small rivers (Fig. 1 and 5, Charente, Lay-Sèvre, Vilaine, Etel, Lorient, Concarneau) could fairly be attributed to a new model of "rocky-coast valley-fill " where hydrodynamics and marine sediment supply are constrained by bedrock morphology. In this model, bedrock control is believed to be of reduced importance in large valley-fill successions (Loire and Gironde) because of the large fluvial sediment input and in the Leyre valley-fill where the underlying bedrock is globally deeper than in other incised-valleys of the Bay of Biscay.

## 6. Conclusions

This study evidences a variability in incised valley-fills that is not shown by previously published models. Because the studied valleys belong to the same basin (the northern and eastern Bay of Biscay), they have experienced similar relative sea-level and climate variations. Hydrodynamics and marine sediment supply variations under the control of bedrock morphology seem to play a key-role by governing the variations in valley-fills from mud-dominated to sand-dominated valley-fills. A "rocky-coast valley-fill" model is proposed for discontinuous and incomplete, valley-fills connected to the small rivers of the northern and eastern Bay of Biscay. This model that implies low subsidence rate, large and frequent sea-level variations, and small continental sediment supply could be extended to other sediment starved margins.

## Acknowledgments

Geosciences-Rennes seismic surveys were funded by the CNRS and Brittany Regional Council (COTARMOR research project, Chief Scientist: JN Proust). CLDG seismic surveys and vibrocore cruise (Chief Scientist: Eric Chaumillon) were funded by the CNRS and Charente-Maritime Council. We thank Th. Garland (EPSHOM) for the submarine bedrock contour map (Fig. 1), P. Guennoc (BRGM) for the large scale fault map (Fig. 1) and X. Bertin (LNEC) for the Isopach map of the sedimentary cover within the Lay-Sèvre, Charente and Seudre incised-valleys (Fig. 4).

Table 1 – River parameters for the main rivers (Drainage basin area > 500 km$^2$) flowing into the Bay of Biscay and connected to the incised-valleys discussed in this study (Gleizes, 1977).

| Rivers (from north to south, Fig. 1) | Drainage Basin Area (km²) | Mean Water discharge m³/s |
|---|---|---|
| Blavet | 620 | 10.3 |
| Vilaine | 4096 | 27.1 |
| Loire | 111000 | 825.0 |
| Lay-Sèvre Niortaise | 936 | 10.7 |
| Charente | 4630 | 58.0 |
| Dordogne | 13800 | 283.0 |
| Garonne | 52000 | 545.0 |

Table 2 – List of recent seismic cruises dedicated to incised valleys located within the inner shelf of the northern and eastern Bay of Biscay.

| Investigated area | Cruise Name and dates | Source frequency | Total length of seismic profiles | References |
|---|---|---|---|---|
| Gironde incised-valley | PLABAS, May 1994 PLACETA, June 1995 | Central frequency = 800 Hz | 200 km | Lericolais et al., 2001 |
| Vilaine, Artimon, Etel, Lorient, Concarneau incised-valleys | GEOVIL, 1998 GEOBLAVET, 2001 GEODET, 2000 | Frequency band = 600-1000 Hz | 3000 km | Proust et al., 2001 Menier, 2004 Menier et al, 2006 |
| Charente and Lay-Sèvre incised-valleys | MOBIDYC1, March 2000, MOBYDYC2, March 2001 MSTULR1, April 2001 | Frequency band = 200-1200 Hz | 3200 km | Weber et al., 2004a Weber et al., 2004b Chaumillon and Weber, 2006 |
| Leyre incised-valley | PLACETA, June 1995 | Central frequency = 800 Hz | 10 km | Féniès and Lericolais, 2005 |
| Loire incised valley | GEOLOIRE July 2004 | Frequency band = 600-1000 Hz | 1100 km | Chaumillon and Proust, 2005 |

Table 3 – Correspondence between the names of the seismic units and boundaries used in this study and in previously published articles used for this synthesis. (IVF = Incised Valley Fill, SB = Surface boundary, TS = Transgressive Surface, TRS : Tidal Ravinement Surface, WRS = Wave Ravinement Surface, BRS = Bay Ravinement Surface, LST = Lowstand System tract, TST = transgressive System Tract, HST = Highstand System tract).

| Seismic units (U) names used in this synthesis, sequence stratigraphy interpretation and global internal geometry | Seismic units and boundaries or reflectors (R) names used in previously published case studies. Already published radiocarbon ages (Before Present) are indicated | | | | | |
|---|---|---|---|---|---|---|
| | Leyre IVF (Féniès and Lericolais, 2005) | Gironde IVF (Lericolais et al., 2001) | Charente IVF (Chaumillon and Weber, 2006) | Lay-Sèvre IVF (Chaumillon and Weber, 2006) | Vilaine IVF (Proust et al., 2001) | Vilaine, Concarneau and Lorient IVF, Menier et al., 2006 |
| | | | *WRS offshore* | | | |
| **U4 (late TST and/or HST)** Low angle to subhorizontal reflectors with the exception of sandbanks | Sand drape not defined as a seismic unit | **S4** (HST) | $U_{IOS}4$ (late TST), $U_{AID}2-4$ *860+/-25 BP* *2205 +/- 30* | $U_{BOS}3$ (late TST),, $U_{BID}2$ (HST) *1230 +/- 30 BP* | **Unit 6** (HST) *8110+/-200 BP* | **Unit** 5 (HST) *8110+/-200 BP* |
| | *R2 = WRS* | *R4 = WRS* | *WRS -TRS* | *WRS-TRS* | *SRU6 = WRS* | *WRS or BRS* |
| **U3 (TST and/or early HST)** High angle reflectors | **US1** (TST) | **S3** (TST) | $U_{IOS}3$ (late TST), | $U_{BID}1$ (late TST) *2815 ± 30 BP* *5260 +/- 30 BP* | **Unit 6** (TST) | **Unit 4** (TST) |
| | | *R3 = TRS* | *TRS* | | *TRS* | *TRS* |
| **U2 (TST)** Low angle reflectors | | **S2** (TST) | $U_{IOS}2$ (TST), *6930 ± 40 BP* | | **Unit 5** (LST) | **Unit 3** (TST) |
| | | *R2 = TS* | *TS* | | *SBU5* | *TS* |
| **U1 (LST)** High angle reflectors | | **S1** (LST) | $U_{IOS}1$ (LST), | | **Unit 4** (LST) | **Unit 1** (LST) |

SB2

Figure Captions

Figure 1 – Simplified bathymetric map of the Bay of Biscay showing the contour of the main incised-valleys evidenced thank to high resolution seismic profiling within the inner shelf. The contour map of submarine rocky outcrops (T. Garland, EPSHOM, pers. com.) and the major faults (P. Guennoc, BRGM, pers. com.) has been also indicated.

Figure 2 – Typical striking seismic lines showing the internal architecture of a mud-dominated valley-fill, the Vilaine incised-valley, a mixed sand and mud-dominated valley-fill, the Charente incised-valley and a sand-dominated valley-fill (modified from Weber et al., 2004a), Lay-Sèvre incised-valley (modified from Weber et al., 2004b). Vertical axis of profiles correspond to ms two way travel time.

Figure 3 – Typical striking interpretations of seismic lines showing the internal architecture of the (1) Concarneau incised-valley, (2) Lorient incised-valley, (3) Vilaine incised-valley, (4) Loire incised-valley, (5) Lay-Sèvre incised-valley (modified from Weber et al., 2004b), (6) Charente incised-valley (modified from Weber et al., 2004a), (7) Gironde incised-valley (modified from Lericolais et al., 2001) and (8) Leyre incised-valley (modified from Féniès and Lericolais, 2005). Vertical axis of profiles correspond to ms two way travel time.

Figure 4 – Bathymetric map (A), position plan of seismic profiles used for the isopach and structural maps (seismic profiles shown in this article are in bold) (B), isopach map of soft sediments overlying Mesozoic bedrock (C) and Structural contour map of the mesozoic bedrock (D) for the Lay-Sèvre, Charente and Seudre incised-valleys (modified from Bertin, 2005). The isopach map (C) clearly shows disconnected depocenters of sediments filling the three valleys.

Figure 5 - Schematic incised-valley cross sections and logs showing the variability of valley-fill within the inner shelf of the Bay of Biscay and its control by the bedrock morphology. SB = Surface Boundary, TS = Transgressive Surface, TRS = Tidal Ravinement Surface, WRS = Wave Ravinement Surface, LST =

Lowstand System Tract, TST = Transgressive System Tract, HST = Highstand System Tract.

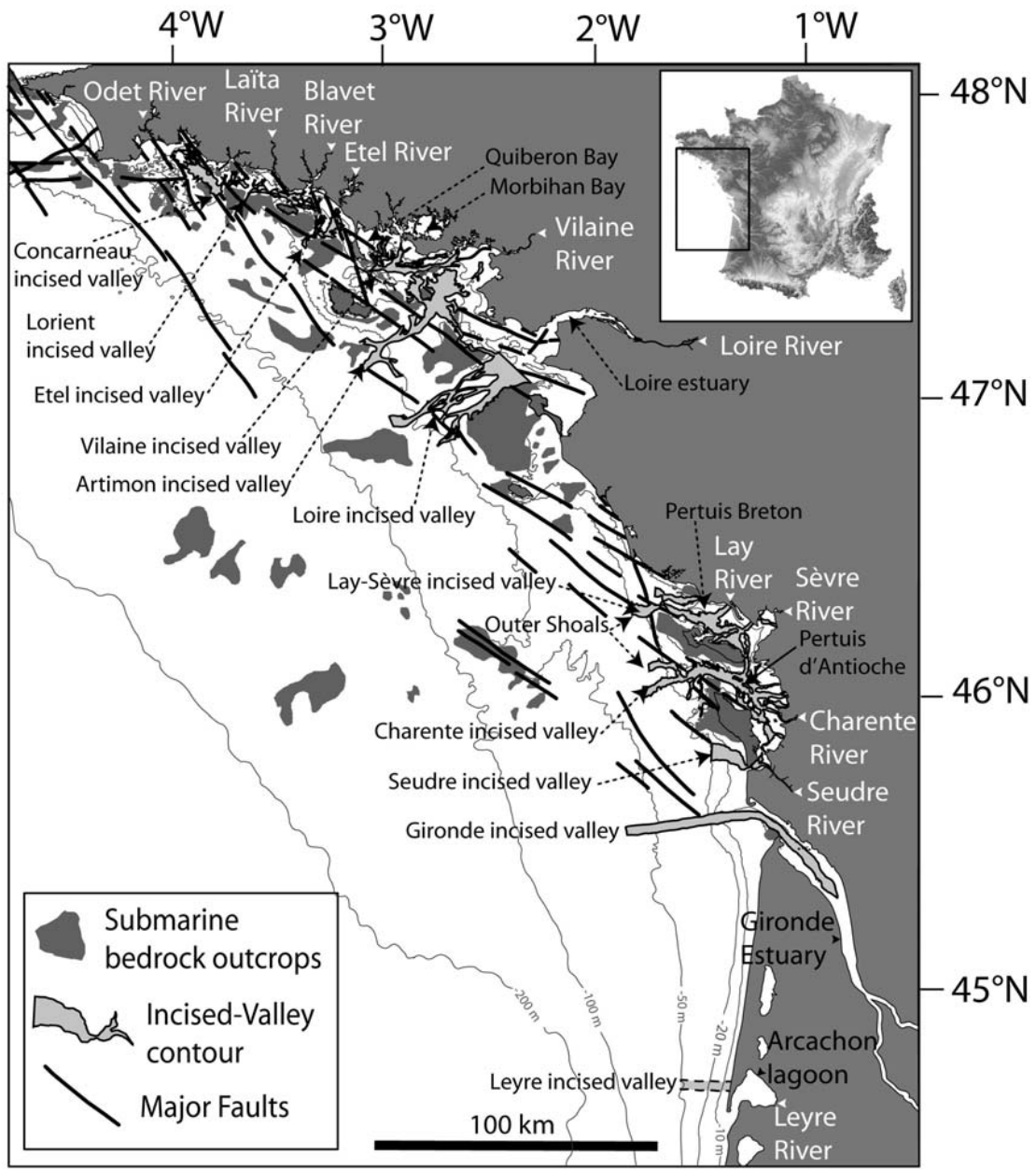

Figure 1

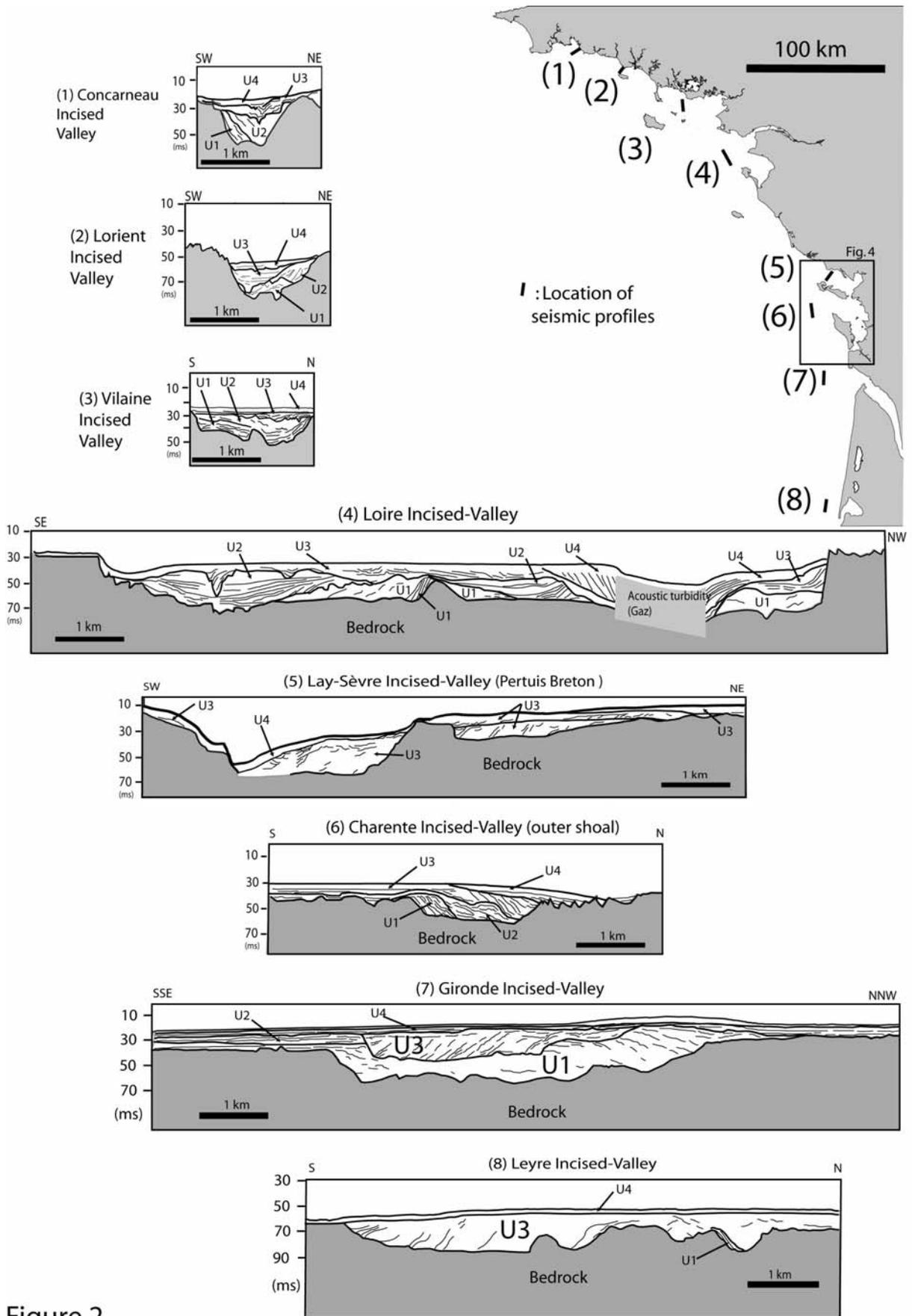

Figure 2

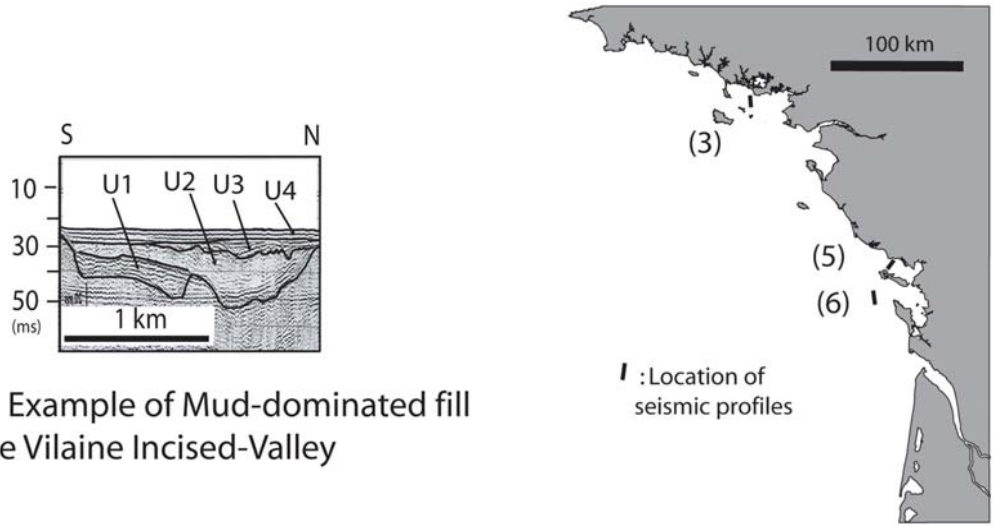

(3) Example of Mud-dominated fill
The Vilaine Incised-Valley

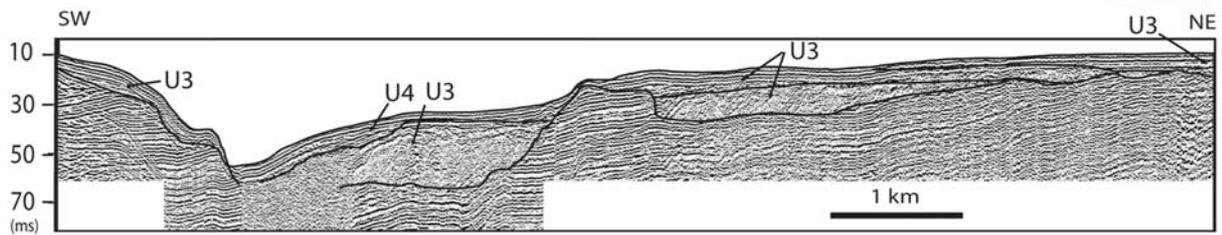

(5) Example of sand-dominated fill
The Lay-Sèvre Incised-Valley (Pertuis Breton)

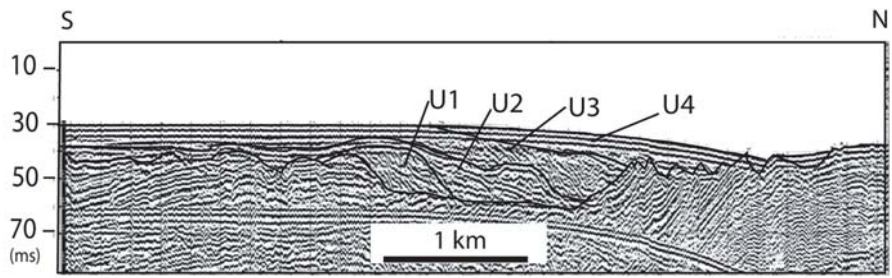

(6) Example of mixed sand and mud dominated fill
The Charente Incised-Valley (outer shoal)

Figure 3

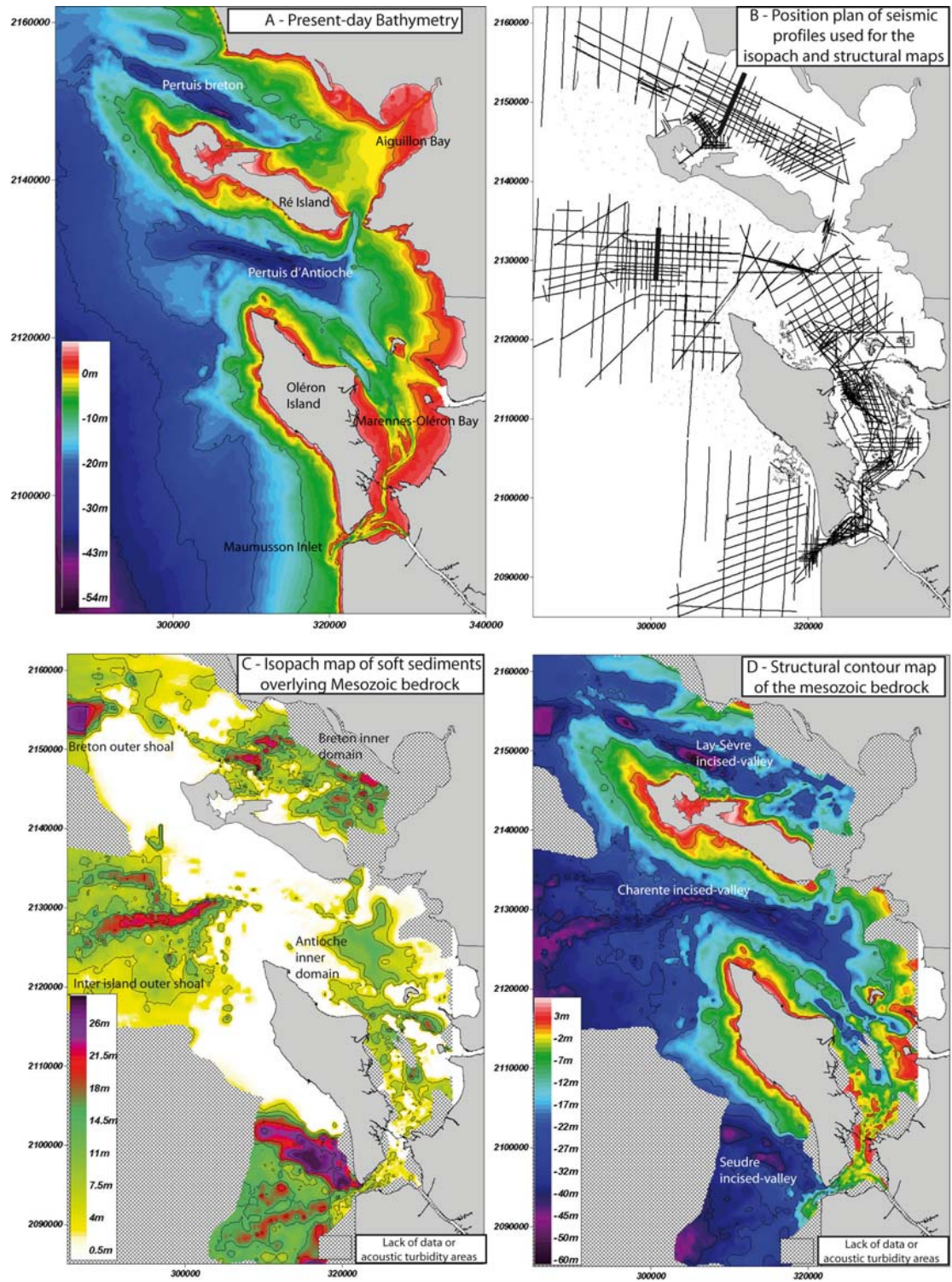

Figure 4

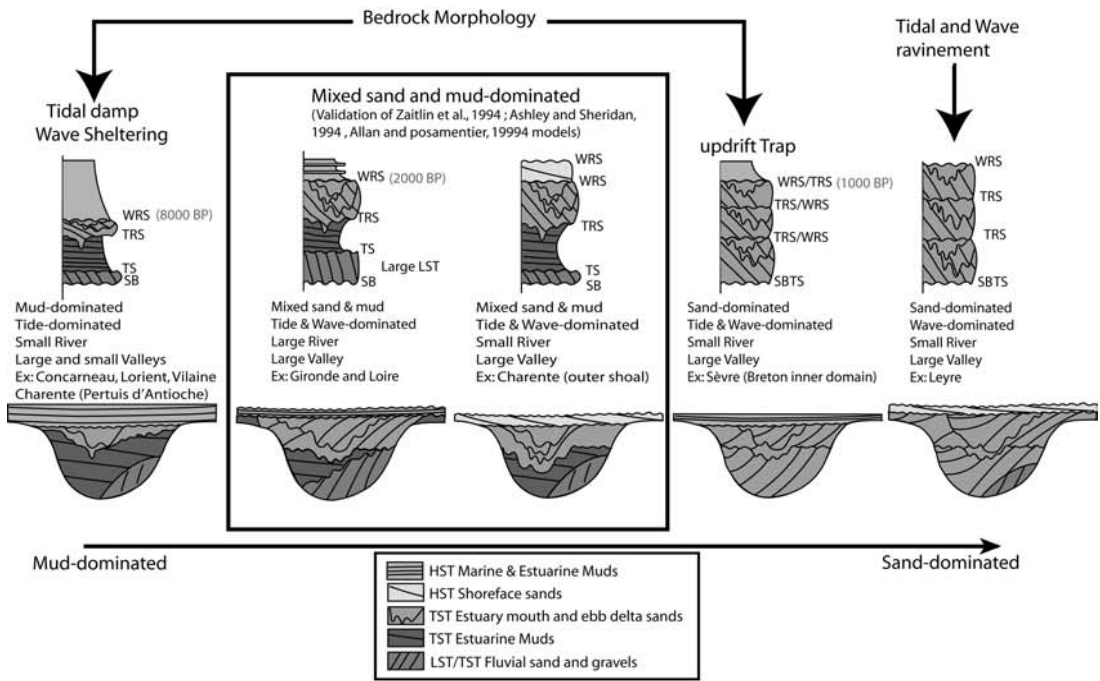

Figure 5